\documentclass[ reprint,superscriptaddress,nofootinbib,amsmath,amssymb,aps]{revtex4-1}
\usepackage{graphicx}
\usepackage{dcolumn}
\usepackage[colorlinks,linkcolor=magenta,anchorcolor=cyan,citecolor=blue]{hyperref}
\usepackage{bm}
\usepackage[multiple]{footmisc}

\def\be{\begin{equation}}
\def\ee{\end{equation}}
\def\ba{\begin{eqnarray}}
\def\ea{\end{eqnarray}}

\def\nn{\nonumber}

\newcommand{\eq}[1]{(\ref{#1})}

\def\nn{\nonumber}\def\q{\theta} \def\r {\rho}     \def\p {\pi} \def\a {\alpha} \def\s {\sigma} \def\d {\delta} \def\f {\phi} \def\g {\gamma}  \def\j {\varphi} \def\k {\kappa} \def\l {\lambda}       \def\pd {\partial}   
\def\Q{\Theta} \def\W{\Omega}             \def\grad{\nabla}\def\.{\cdot}
\def\math {\mathcal}
\begin{document}

\title{Entropy increases at linear order in scalar-hairy Lovelock gravity}
\author{Jie Jiang}
\email{jiejiang@mail.bnu.edu.cn}
\affiliation{College of Physics and Communication Electronics, Jiangxi Normal University, Nanchang 330022, China\label{addr2}}
\affiliation{Department of Physics, Beijing Normal University, Beijing 100875, China\label{addr2}}
\author{Ming Zhang}
\email{Corresponding author. mingzhang@jxnu.edu.cn}
\affiliation{College of Physics and Communication Electronics, Jiangxi Normal University, Nanchang 330022, China\label{addr2}}
\date{\today}

\begin{abstract}
In this paper, we investigate the second law of the black holes in Lovelock gravity sourced by a conformally coupled scalar field under the first-order approximation when the perturbation matter fields satisfy the null energy condition. First of all, we show that the Wald entropy of this theory does not obey the linearized second law for the scalar-hairy Lovelock gravity which contains the higher curvature terms even if we replace the gravitational part of Wald entropy with Jacobson-Myers (JM) entropy. This implies that we cannot naively add the scalar field term of the Wald entropy to the JM entropy of the purely Lovelock gravity to get a valid linearized second law. By rescaling the metric, the action of the scalar field can be written as a purely Lovelock action with another metric. Using this property, by analogy with the JM entropy of the purely Lovelock gravity, we introduce a new formula of the entropy in the scalar-hairy Lovelock gravity. Then, we show that this new JM entropy increases along the event horizon for Vaidya-like black hole solutions and therefore it obeys a linearized second law. Moreover, we show that different from the entropy in $F($Riemann$)$ gravity, the difference between the JM entropy and Wald entropy also contains some additional corrections from the scalar field.

\end{abstract}
\maketitle
\section{Introduction}

In $1970$, Bekenstein, Hawking, Davies, and Unruh showed separately that the dynamical quantities of the horizon in general relativity can be treated as the thermodynamical variables \cite{A1,A2,A3,A4,A5,A6}. Therefore, the laws of the thermodynamical system should also be held for the black hole system. The most profound laws are the first and second laws of black hole mechanics. In general relativity, the first law shows that $dM=TdS$ for the stationary black hole where $M$ is the energy of the black hole, the temperature $T$ is proportional to the surface gravity $\k$, and entropy $S$ is proportional to the area $A$ of the cross-section on the event horizon. Then, the second law becomes that the area $A$ of the horizon increases irreversibly \cite{A1,A7,A8}. Also, the generalized second law states that the sum of the entropies of the horizon and the matter outside is always increasing \cite{A9,A10,A11,A12}.

After the quantum effect or string modification is taken into account, higher curvature term should be added to the Einstein-Hilbert action \cite{A13,A14,A15,A16}. A natural question is to ask whether the black hole in any generally covariant gravitational theory can be regarded as a thermodynamical system. Therefore, Wald and collaborators obtained the first law of the stationary black holes for any diffeomorphism invariant gravitational theory based on the Noether charge method \cite{A17,A18}. In their results, the entropy of the stationary black hole is a local geometry quantity which is integrated over a spacelike cross-section $s$ on the horizon, i.e.,
\ba\begin{aligned}
S_W=-2\p\int_s d^{D-2}x\sqrt{\g}\frac{\d \math{L}}{\d R_{abcd}}\hat{\bm{\epsilon}}_{ab}\hat{\bm{\epsilon}}_{cd}\,,
\end{aligned}\ea
where $\math{L}$ is the Lagrangian density of the gravitational theory, $\hat{\bm{\epsilon}}_{ab}$ is the binormal of the cross-section $s$, and $\g_{ab}$ is the intrinsic metric of $s$. The ``physical process version'' of the first law was investigated in Refs. \cite{A19,A20,A21}.
However, there exists some ambiguity when we consider a nonstationary black hole and the Wald entropy is just one of the possible candidates for entropy \cite{A18,A21}. All of these candidates are only off by some quantities which are vanishing for stationary cases.

Therefore, the most important thing is to see whether the higher curvature corrections spoil the second law of black hole thermodynamics and the entropy satisfies the second law of black hole thermodynamics. In Refs. \cite{A21,A22,A23}, based on the field redefinition, it was shown that the Wald entropy obeys the second law for the $f(R)$ gravity. For other cases with higher curvature terms, there are some violations of the second law when two black holes merge \cite{A24}. However, as mentioned in \cite{Bhattacharjee:2015yaa}, if we want to consider the general second law in the case of an adiabatic change in a quantum black hole, it is enough for us to consider it at linear order. If we restrict attention to linearized metric perturbations to stationary black holes, it has been shown that the Jacobson-Myers (JM) entropy of the Lovelock gravity and the holographic entropy of quadratic curvature gravity obey the second law \cite{Bhattacharjee:2015yaa,A25,A26,A27}. More generally, Wall gave a general method to evaluate the corrected entropy which satisfies the linearized second law and showed that it takes the form \cite{Wall}
\ba\begin{aligned}\label{df1}
S&=-2\p\int_s d^{D-2}x\sqrt{\g}\left[\frac{\pd F}{\pd R_{abcd}}\hat{\bm{\epsilon}}_{ab}\hat{\bm{\epsilon}}_{cd}\right.\\
&\left.+8\frac{\pd^2 F}{\pd R_{k a i b}\pd R^k{}_{c j d}}K^{(i)}_{ab}K^{(j)}_{cd}\right]
\end{aligned}\ea
at linear order in $F($Riemann$)$ gravity, in which $K^{(i)}_{ab}$ is the extrinsic curvature corresponding to the normal direction of the cross-section $s$, and $i, j, k$ denote the indices of the orthogonal vectors in the two-normal vector space.

Most recently, it has been shown that Lovelock gravity can be conformally coupled to a scalar field \cite{B1}, which is called scalar-hairy Lovelock gravity in the following discussion. This theory admits a black hole solution with conformal scalar hair, i.e., the scalar field is nonvanishing and regular everywhere outside of the singularity \cite{B2, B3, B4}. In Ref. \cite{Hennigar:2016ekz}, the thermodynamics of the stationary black hole in this theory has been investigated. They showed that the Wald entropy satisfies the first law.  Because the Lagrangian of the scalar field contains the Riemann curvature and the scalar field is nonvanishing, the Wald entropy should also be corrected by the scalar field even for the stationary black hole cases. However, as we all know, the JM entropy of the purely Lovelock gravity obeys the linearized second law but that Wald entropy does not \cite{A25, A26, A27}. Therefore, as an extension of purely Lovelock gravity, it is not difficult to believe that proper entropy of the scalar-hairy Lovelock gravity should also be amended. Can we naively add the scalar field term of the Wald entropy to the JM entropy of the purely Lovelock gravity to get a valid linearized second law in this theory? If it is not, how to construct a corrected one?

The outline of this paper is as follows. In Sec. \ref{sec2}, we review the gravitational theory containing a real scalar field $\f$ conformally coupled to Lovelock gravity. In Sec. \ref{sec3}, we turn to examine the linearized second law in the Vaidya-like black hole solution for scalar-hairy Lovelock gravity. To be specific, in Sec. \ref{sec31}, we evaluate the increases of the Wald entropy along the event horizon. In Sec. \ref{sec32}, based on the property that the action of the scalar field can be expressed as a purely Lovelock term after a conformal transformation, we construct a JM entropy for the scalar-hairy Lovelock gravity by analogy with that of purely Lovelock case and then check its corresponding linearized second law in the Vaidya-like solution. Finally, the conclusions are presented in Sec. \ref{sec5}.

\section{scalar-hairy Lovelock gravity}\label{sec2}

In this paper, we consider a gravitational theory containing a real scalar field $\f$ conformally coupled to Lovelock gravity. The action of this theory in $D$-dimensional spacetime is given by \cite{B1}
\ba\begin{aligned}
I=\frac{1}{16\p}\int d^Dx\sqrt{g}\left(\sum_{k=0}^{k_\text{max}}\math{L}^{(k)}+\math{L}_\text{mat}\right),
\end{aligned}\ea
where $\math{L}_\text{mat}$ is the Lagrangian density of the extra matter fields, and
\ba\begin{aligned}
\math{L}^{(k)}=\frac{1}{2^k}\d^{(k)}\left(a_k R^{(k)}+b_k\f^{d-4k}S^{(k)}\right)
\end{aligned}\ea
is the $k$-order Lagrangian density with some parameters $a_k$ and $b_k$. Here $k_\text{max}=\left[(D-1)/2\right]$ and we have denoted
\ba\begin{aligned}
R^{(k)}=\prod_{r=1}^k R^{c_r d_r}_{a_rb_r}\,,\ \ \ \ S^{(k)}=\prod_{r=1}^k S^{c_r d_r}_{a_rb_r}
\end{aligned}\ea
with the generalized Kronecker tensor
\ba\begin{aligned}
\d^{(k)}=(2k)!\d^{[a_1}_{c_1}\d^{b_1}_{d_1}\cdots\d^{a_k}_{c_k}\d^{b_k]}_{d_k}\,,
\end{aligned}\ea
in which $R^{cd}_{ab}$ is the Riemann tensor of the metric $g_{ab}$ and $S^{cd}_{ab}$ is defined as
\ba\begin{aligned}
S^{cd}_{ab}&=\f^2 R^{cd}_{ab}-2\d_{[a}^{[c}\d_{b]}^{d]}\grad_e\f\grad^e\f\\
&-4\f \d_{[a}^{[c}\grad_{b]}\grad^{d]}\f+8\d_{[a}^{[c}\grad_{b]}\f\grad^{d]}\f\,.
\end{aligned}\ea
The equations of motion derived from varying this action read
\ba\begin{aligned}
&G_{ab}=T^\f_{ab}+8\p T_{ab}\,,\\
&\sum^{k_\text{max}}_{k=0}\frac{(D-2k)b_k}{2^k}\f^{D-4k-1}\d^{(k)}S^{(k)}=\j\,,
\end{aligned}\ea
with the generalized Einstein tensor
\ba\begin{aligned}
G_a^b=-\sum^{k_\text{max}}_{k=0}\frac{a_k}{2^{k+1}}\d_{ac_1d_1\cdots c_kd_k}^{b a_1b_1\cdots a_k b_k}R^{c_1d_1}_{a_1b_1}\cdots R^{c_kd_k}_{a_kb_k}
\end{aligned}\ea
and the stress-energy tensor of the scalar field
\ba\begin{aligned}
(T^\f)_a^b&=\sum^{k_\text{max}}_{k=0}\f^{D-4k}\frac{a_k}{2^{k+1}}\d_{ac_1d_1\cdots c_kd_k}^{b a_1b_1\cdots a_k d_k}S^{c_1d_1}_{a_1b_1}\cdots S^{c_kd_k}_{a_kb_k}\,.
\end{aligned}\ea
Here $T_{ab}$ and $\j$ are the stress-energy tensor and source of the perturbation matter fields. It is not difficult to verify that this theory is invariant under the conformal transformation: $g_{ab}\to \W^2 g_{ab}$ and $\f\to \W^{-1}\f$. It can be regarded as a natural generalization of Lovelock gravity with a non-minimal coupling scalar field. This theory admits a scalar-hairy black hole solution where the scalar field is nonvanishing and regular outside of the singularity \cite{B2, B3, B4, Hennigar:2016ekz}.

Employing the Noether charge method \cite{A17,A18}, the Wald entropy of this gravitational theory can be obtained and it is given as
\ba\begin{aligned}
S_W=-2\p\int_s d^{D-2}x\sqrt{\g}P^{abcd}\hat{\bm{\epsilon}}_{ab}\hat{\bm{\epsilon}}_{cd}\,,
\end{aligned}\ea
where we have denoted
\ba\begin{aligned}
P^{abcd}=\sum_{k=0}^{k_\text{max}}\frac{\pd \math{L}^{(k)}}{\pd R_{abcd}},
\end{aligned}\ea
and it can be expressed as
\ba\begin{aligned}
P_{ab}^{cd}=\frac{1}{16\p}\sum^{k_\text{max}}_{k=0}\frac{k(2k)!}{2^k}\d_{[a}^c\d_b^d\d_{a_2}^{c_2}\d_{b_2}^{d_2}\cdots\d_{a_k}^{c_k}\d_{{b_k]}}^{d_k}\\
\times \left[ a_k \prod_{r=2}^kR^{a_rb_r}_{c_r d_r}+b_k \f^{D-4k+2}\prod_{r=2}^kS^{a_rb_r}_{c_r d_r}\right]\,.
\end{aligned}\ea
Here $s$ is a cross-section of event horizon, $\g_{ab}$ is the induced metric on $s$, and $\hat{\bm{\epsilon}}_{ab}$ is the binormal to $s$. Because the scalar field can be nonvanishing, the Wald entropy in this theory should also be corrected by the scalar field even for the stationary black hole cases. The validity of the first law and thermodynamics of this entropy in charged scalar-hairy black holes have been discussed in \cite{Hennigar:2016ekz}. However, as mentioned in the introduction, the Wald entropy of the purely Lovelock gravity does not obey the linearized second law and we need to focus on the JM entropy \cite{A25, A26, A27}. As a direct extension of the purely Lovelock gravity, it is natural for us to replace the gravitational part of Wald entropy with JM entropy in the scalar-hairy Lovelock gravity. What about the scalar field part? Can we directly utilize the scalar field term of the Wald entropy? In the following, we would like to investigate these questions by examining Vaidya-like solutions in the scalar-hairy Lovelock gravity. Since the case with $a_k=b_k=0$ for $k\geq 2$ is equivalent to the Einstein gravity minimally coupled to the scalar field after performing a field redefinition and it obeys the second law, we next only consider the case which contains the higher curvature terms.

\section{linearized second law of the scalar-hairy Lovelock gravity}\label{sec3}

The main purpose of this paper is to check whether the entropy is increasing along the event horizon in the physical process under the linear order of perturbation. Therefore, we need to assume that spacetime is a black hole, i.e., the physical process satisfies the weak cosmic censorship conjecture \cite{RPenrose}. With similar consideration of \cite{Bhattacharjee:2015yaa}, in the following, we would like to test the linearized second law in a Vaidya solution which can be constructed by the static spherically symmetric black hole with infalling matter source. Without loss of generality, the line element can be expressed as the form
\ba\begin{aligned}\label{ds2}
ds^2=-f(r,v)dv^2+2dv dr+r^2 d\W_{D-2}^2\,,
\end{aligned}\ea
where $f(r, v)$ is an arbitrary function. The horizon of this black hole is located at $r=r(v)$ which is obtained from $r'(v)=f(r,v)/2$ with some appropriate boundary conditions. The null generator of the horizon with the nonaffine parameter $\l$ is given by
\ba\begin{aligned}\label{0422}
k^a=\left(\frac{\pd}{\pd \l}\right)^a=2\left(\frac{\pd}{\pd v}\right)^a+f(r, v)\left(\frac{\pd}{\pd r}\right)^a\,.
\end{aligned}\ea
The second null vector on the event horizon which satisfies $k^a l_a=-1$ is expressed as
\ba\begin{aligned}
l_a=-\frac{1}{2}(dv)_a\,.
\end{aligned}\ea
Then, we have $k^a\grad_a k^b=\k k^b$ with $\k=f'(r,v)$. In this paper, we would like to consider the situation when a static black hole is perturbed by some extra matter fields and ultimately settle down to a static state in the asymptotic future. This implies that $f(r, v)$ will be independent on $v$ and $k^a$ is an exact Killing vector at sufficiently late times. For the background geometry, $\k/2$ will be the surface gravity of the event horizon. Since we turn to test the second law under the first-order approximation of perturbation, we introduce a small parameter $\a$ such that $f(r, v)=f(r)+\a \d f(r, v)$ at first-order of $\a$. Then, one can check that $\q\sim \s_{ab}\sim \a$ with the expansion $\q$ and shear $\sigma_{ab}$ of the event horizon. Then, the Raychaudhuri equation gives
\ba\begin{aligned}
\frac{d\q}{d\l}&=-\frac{\q^2}{D-2}-\s_{ab}\s^{ab}-R_{ab}k^ak^b+\k \q\\
&\simeq -R_{kk}+\k \q
\end{aligned}\ea
under the first-order approximation of $\a$. Next, we define the entropy density $\r$ as
\ba\begin{aligned}
S=\frac{1}{4}\int_s d^{D-2}x \sqrt{\g} \r\,.
\end{aligned}\ea
Then, we can define the change of entropy per unit area as a generalized expansion given by
\ba\begin{aligned}
\frac{d S}{d\l}=\frac{1}{4}\int_s d^{D-2}x \sqrt{\g} \Q\,.
\end{aligned}\ea
With simple calculation, we have
\ba\begin{aligned}
\Q=\frac{d \r}{d\l}+\q \r\,.
\end{aligned}\ea
By calculating the change of $\Q$, we can easily obtain
\ba\begin{aligned}\label{QTE}
\frac{d\Q}{d \l}-\k \Q=-8\p T_{kk}+E_{kk}
\end{aligned}\ea
with
\ba\begin{aligned}
E_{kk}= G_{kk}-T^\f_{kk}+\grad_k \grad_k \r-\r R_{kk}\,,
\end{aligned}\ea
in which we have denoted $A_{kk}=A_{ab}k^ak^b$ for the tensor $A_{ab}$.

According to the discussion in \cite{Bhattacharjee:2015yaa}, the key point to prove the entropy increases along the event horizon is to show that $E_{kk}\simeq 0$ under the first-order approximation. In the following, we turn to review this result. If $E_{kk}\simeq 0$ at first order, combing the null energy condition $T_{kk}\geq 0$, Eq. \eq{QTE} reduces to
\ba\begin{aligned}
\frac{d\Q}{d\l}-\k \Q\leq 0\,.
\end{aligned}\ea
Under the assumption that the black hole becomes a static state in the asymptotic future, we have $\Q=0$ at late times. When we focus on the calculation under first order, the surface gravity $\k$ in \eq{QTE} is evaluated at zero-order and it should be positive. Then, it is easy to verify that $\Q$ must be positive everwhere on the event horizon, i.e., the entropy satisfies the linearized second law. Moreover, it has been shown in \cite{A27} that this condition also gives the generalized second law in linear order. For simplification, unless otherwise specified, the rest of the calculations are done in the first order.

Using the explicit expression of metric in \eq{ds2}, we can further obtain
\ba\begin{aligned}\label{EGT}
E_{kk}&\simeq G_{kk}-T^\f_{kk}-\r R_{kk}\\
&+4\pd_v^2 \r-2f'(r) \pd_v\r+2\r'(r) \pd_v f\,.
\end{aligned}\ea
Performing the explicit expression of line element in \eq{ds2}, we give some useful quantities as follows
\ba\begin{aligned}
y&=R_{ai}^{bi}k^ak_b=-\frac{2 \pd_v f}{r}\,,\ \ \ x=R_{ij}^{ij}=\frac{1-f}{r^2}\,,\\
\tilde{y}&=S_{ai}^{bi}k^ak_b=\f^2y+2\f(\pd_v\f f'-2\pd_v^2\f-\f'\pd_vf)\,,\\
\tilde{x}&=S_{ij}^{ij}=\f^2x-\frac{2\f(\pd_v\f+f \f')}{r}-\f'(2\pd_v\f+f \f')\,.
\end{aligned}\ea
Then, the first two term of Eq. \eq{EGT} can be obtained and it can be expressed as
\ba\begin{aligned}
&G_{kk}-T^\f_{kk}\\
&=\sum_{k=0}^{k_\text{max}}\left(\frac{k(D-2)!}{(D-2k-1)!}\frac{a_ky+b_k\f^{D-2k-2}\tilde{y}}{r^{2k-2}}\right)\,,
\end{aligned}\ea
For the third term, because $R_{kk}=(D-2)y$ is the first-order quantity, the entropy $\r$ should be evaluated at zero-order. Since all of the candidates are same in the stationary case, we can directly use the expression of Wald entropy to evaluate it. Under the background geometry, we have
\ba\begin{aligned}
\r\simeq \r_W\simeq \sum_{k=1}^{k_\text{max}}\left(\frac{k(D-2)!}{(D-2k)!}\frac{a_k+b_k\f^{D-2k}}{r^{2k-2}}\right)\,,
\end{aligned}\ea
where the density of the Wald entropy is defined by
\ba\begin{aligned}\label{Walddensity}
\r_{W}=-8\p P^{abcd}\hat{\bm{\epsilon}}_{ab}\hat{\bm{\epsilon}}_{cd}\,.
\end{aligned}\ea
Combing above results, the first three terms can be expressed as
\ba\begin{aligned}
&G_{kk}-T^\f_{kk}-\r R_{kk}=\sum_{k=1}^{k_\text{max}}\left(\frac{2k(D-2)!}{r^{2k-2}(D-2k)!}\right)\\
&\times\left[(1-k)(a_k+b_k\f^{D-2k})y\right.\\
&\left.+b_k(D-2k)\f^{D-2k-1}(f'\pd_v\f-2\pd_v^2\f-\f'\pd_vf)\right]\,.
\end{aligned}\ea

\subsection{Wald entropy}\label{sec31}
In this subsection, we start by considering the Wald entropy as shown in the last section. Performing the explicit expression of line element in \eq{ds2}, the Wald entropy density can be further obtained and it can be expressed as
\ba\begin{aligned}
\r_W=\sum_{k=1}^{k_\text{max}}\left[\frac{k(D-2)!}{(D-2k)!}\left(a_kx^{k-1}+b_k\f^{D-4k+2}\tilde{x}^{k-1}\right)\right]\,.
\end{aligned}\nn\\\ea

For the second line of \eq{EGT}, we have
\ba\begin{aligned}
&4\pd_v^2 \r_W-2f' \pd_v\r_W+2\r'_W \pd_v f=\sum_{k=0}^{k_\text{max}}\left(\frac{2k(D-2)!}{(D-2k)!}\right)\\
&\times\left[\frac{(a_k+b_k\f^{D-2k})}{r^{2k-2}}(k-1)(y+f'\pd_vf-2\pd_v^2f)\right.\\
&\left.-\frac{b_k\f^{D-2k-1}}{r^{2k-2}}(D-2k)(f'\pd_v\f-2\pd_v^2\f-\f'\pd_vf)\right.\\
&\left.+\frac{b_k\f^{D-2k-2}}{r^{2k-4}}(k-1)\left(\frac{4\f(\pd_v^3\f+\f'\pd_v^2f)}{r}\right.\right.\\
&+2\f'(2\pd_v^3\f+\pd_v^2f \f')-f'\f'(2\pd_v^2\f+ \f'\pd_vf)\\
&\left.\left.-\frac{2\f f'(\pd_v^2\f+\f'\pd_vf)}{r}\right)\right]\,.
\end{aligned}\ea
Combing above results, we can obtain
\ba\begin{aligned}
&E_{kk}=\sum_{k=2}^{k_\text{max}}\left(\frac{2k(k-1)(D-2)!}{(D-2k)!}\right)\\
&\times\left[\frac{(a_k+b_k\f^{D-2k})}{r^{2k-2}}(f'\pd_vf-2\pd_v^2f)\right.\\
&\left.+\frac{b_k\f^{D-2k-2}}{r^{2k-4}}\left(\frac{4\f(\pd_v^3\f+\f'\pd_v^2f)}{r}\right.\right.\\
&+2\f'(2\pd_v^3\f+\pd_v^2f \f')-f'\f'(2\pd_v^2\f+ \f'\pd_vf)\\
&\left.\left.-\frac{2\f f'(\pd_v^2\f+\f'\pd_vf)}{r}\right)\right]\,.
\end{aligned}\ea

As mentioned in the last section, we focus on the case where the action contains the higher curvature terms, i.e., there are at least one non-zero parameters $a_k$ or $b_k$ for $k\geq 2$. Then, $E_{kk}$ will be nonvanishing. Since the null energy condition for the scalar field $T_{kk}\geq 0$ only depends on the first two order derivative of $\f$ and the first derivative of $f$, there does not exist any constraints on $\pd_v^2 f$ and $\pd_v^3\f$, which indicates that $E_{kk}$ need not to have any specific sign. Therefore, the linearized second law for the Wald entropy is violated in the scalar-hairy Lovelock gravity.

Next, we consider the entropy after replacing the gravitational part of Wald entropy with JM entropy in the scalar-hairy Lovelock gravity. It is not difficult to believe that this correction will only change the quantities which contain the coefficient  $a_k$. This implies that there exists at least some non-zero term containing $\pd_v^3\f$. Therefore, the linearized second law for this corrected entropy is also violated and we cannot naively add the scalar field term of the Wald entropy to the JM entropy of the purely Lovelock gravity.

\subsection{Jocobson-Myers entropy}\label{sec32}

In this subsection, we turn to construct a horizon entropy such that the linearized second law is satisfied. Note that the theory is conformally invariant. By rescaling the metric $\tilde{g}_{ab}=\f^2 g_{ab}$, we can find
\ba\begin{aligned}
\tilde{R}^{cd}_{ab}=\f^{-4} S^{cd}_{ab}\,,
\end{aligned}\ea
where $\tilde{R}^{cd}_{ab}$ are associated to the metric $\tilde{g}_{ab}$ and all of the indexes are raised by $\tilde{g}^{ab}$ for the quantities with $\sim$. Then, the action of the scalar-hairy Lovelock gravity can be expressed as
\ba\begin{aligned}\label{III}
I=I^{\{a\}}[g]+I^{(b)}[\tilde{g}]\,,
\end{aligned}\ea
where $I^{\{c\}}[g]$ is the action of the purely Lovelock gravity with the metric $g_{ab}$ and coupling parameter $\{c_1, c_2, \cdots c_{k_\text{max}}\}$, i.e.,
\ba\begin{aligned}\label{Apure}
I^{\{c\}}[g]=\frac{1}{16\p}\sum_{k=0}^{k_\text{max}}\left[\int_s d^Dx \sqrt{-g} \frac{c_k}{2^k}\d^{(k)}R^{(k)}\right]\,.
\end{aligned}\ea

As mentioned in the section of the introduction, it is the JM entropy of purely Lovelock gravity that satisfies the linearized second law, not the Wald entropy. By analogy with the JM entropy of the purely Lovelock gravity, we can also introduce a JM entropy in the scalar-hairy Lovelock gravity as
\ba\begin{aligned}
S_\text{JM}=S_\text{JM}^{\{a\}}[g]+S^{(b)}_\text{JM}[\tilde{g}]\,,
\end{aligned}\ea
where $S_\text{JM}^{\{c\}}[g]$ is the JM entropy of the purely Lovelock gravity with the coefficients $\{c_k\}$ and metric $g_{ab}$, and it can be shown as
\ba\begin{aligned}
S_\text{JM}^{\{c\}}[g]=\frac{1}{4}\sum^{k_\text{max}}_{k=0}\left[\frac{kc_k}{2^k}\int_s d^{D-2}x\sqrt{\g}\d^{(k-1)}\math{R}^{(k-1)}\right]_g\,.\,.
\end{aligned}\ea
where the subscript $g$ means that all of the quantities are evaluated on the metric $g_{ab}$, and we have denoted
\ba\begin{aligned}
\math{R}^{(k)}=\prod^{k}_{r=1}\math{R}_{a_rb_r}^{c_rd_r}
\end{aligned}\ea
with the intrinsic curvature $\math{R}_{ab}^{cd}$ of the cross-section $s$ on the horizon. Then, the JM entropy density can be further obtained
\ba\begin{aligned}
\r_\text{JM}=\sum^\text{max}_{k=0}\left[\frac{k}{2^k}\d^{(k-1)}\left(a_k \math{R}^{(k-1)}+b_k\f^{D-2}\tilde{\math{R}}^{(k-1)}\right)\right].
\end{aligned}\nn\\\ea
We can see that this entropy can reduce to the Wald entropy in the static black hole geometry. From the discussion in \cite{Wall}, the difference between the Wald entropy and JM entropy of the purely Lovelock gravity is given by
\ba\begin{aligned}
&S_W^{\{a\}}[g]-S_\text{JM}^{\{a\}}[g]\\
&=16\p\int_s d^{D-2}x\sqrt{\g}\frac{\pd^2 \math{L}^{\{a\}}[g]}{\pd R_{k a i b}\pd R^k{}_{c j d}}K^{(i)}_{ab}K^{(j)}_{cd}
\end{aligned}\ea
at linear order. Here $\math{L}^{(a)}[g]$ is the Lagrangian density of the purely Lovelock gravity with action $I^{(a)}[g]$ in \eq{Apure} the extrinsic curvature of the cross-section is given by
\ba\begin{aligned}
K^{(i)}_{ab}=\frac{1}{2}\math{L}_{n^{(i)}}\g_{ab}\,,
\end{aligned}\ea
and the orthogonal normal vectors $n_a^{(i)}$ with $i=1,2$ are defined as
\ba\begin{aligned}
n^{(1)}_a&=\sqrt{|f|}\left[(dv)_a-\frac{1}{f}(dr)_a\right]\,,\\
n^{(2)}_a&=\frac{1}{\sqrt{|f|}}(dr)_a\,.
\end{aligned}\ea

The above result can naturally give the deference of the JM entropy and Wald entropy in scalar-hairy Lovelock gravity\ba\begin{aligned}\label{relation1}
&S_W-S_\text{JM}=16\p\int_s d^{D-2}x\sqrt{\g}\frac{\pd^2 \math{L}^{\{a\}}[g]}{\pd R_{k a i b}\pd R^k{}_{c j d}}K^{(i)}_{ab}K^{(j)}_{cd}\\
&+16\p\int_s d^{D-2}x\sqrt{\tilde{\g}}\frac{\pd^2 \math{L}^{\{b\}}[\tilde{g}]}{\pd \tilde{R}_{k a i b}\pd \tilde{R}^k{}_{c j d}}\tilde{K}^{(i)}_{ab}\tilde{K}^{(j)}_{cd}\,,
\end{aligned}\ea
where $\tilde{K}^{(i)}_{ab}$ is the extrinsic curvature corresponding to the normal vectors $\tilde{n}_a^{(i)}=\f n^{(i)}_a$ and evaluated in the spacetime with metric $\tilde{g}_{ab}$. Performing the conformal transformation $\tilde{g}_{ab}=\f^2 g_{ab}$ and $\tilde{\g}_{ab}=\f^2 \g_{ab}$, we can further obtain
\ba\begin{aligned}
\tilde{K}^{(i)}_{ab}=\f K^{(i)}_{ab}+2\g_{ab}\grad^i \f\,,
\end{aligned}\ea
where we have denoted $\grad^i \f =n^{(i)}_a\grad^a\f$. Then, the relationship of these two entropy can be expressed by the quantities in the spacetime with metric $g_{ab}$, i.e.,
\ba\begin{aligned}\label{relation1}
&S_W-S_\text{JM}=16\p\int_s d^{D-2}x\sqrt{\g}\left[\frac{\pd^2 \math{L}}{\pd R_{k a i b}\pd R^k{}_{c j d}}K^{(i)}_{ab}K^{(j)}_{cd}\right.\\
&\left.+4\f^4\frac{\pd^2 \math{L}}{\pd S_{k a i b}\pd S^k{}_{c j d}}\g_{cd}\grad^j\ln\f\left(K^{(i)}_{ab}+\g_{ab}\grad^i\ln\f\right)\right]\,.
\end{aligned}\ea
We can see that the difference between above JM entropy and IW entropy is different from the result of the $F($Riemann$)$ gravity as shown in \eq{df1}, and they also contain some correction from the scalar field.

In the following, we would like to check whether the JM entropy increases along the event horizon at first order in the Vaidya-like black hole solutions if the perturbation matter fields satisfy the null energy condition. Using the line element in \eq{ds2}, we can further obtain
\ba\begin{aligned}
\r_\text{JM}=\sum_{k=0}^{k_\text{max}}\left[\frac{k(D-2)!}{(D-2k)!}\frac{(a_k+b_k\f^{D-2k})}{r^{2k-2}}\right]\,.
\end{aligned}\ea
From the calculation at the beginning of this section, we only need to evaluate the second line of \eq{EGT}. Using the above JM entropy density, we have\ba\begin{aligned}
&4\pd_v^2 \r_\text{JM}-2f' \pd_v\r_\text{JM}+2\r'_\text{JM} \pd_v f=\sum_{k=0}^{k_\text{max}}\left(\frac{2k(D-2)!}{(D-2k)!}\right)\\
&\times\left[(k-1)(a_k+b_k\f^{D-2k})yx^{k-1}\right.\\
&\left.-b_k(D-2k)\f^{D-2k-1}x^{k-1}(f'\pd_v\f-2\pd_v^2\f-\f'\pd_vf)\right]\,.
\end{aligned}\ea
Combing above results, we can see that
\ba
E_{kk}= 0
\ea
under the first-order approximation. This means that the JM entropy increases along the horizon in the Vaidya-like black hole at the first-order approximation of perturbation. This also implies that similar to the JM entropy in purely Lovelock gravity, the JM entropy in the scalar-hair Lovelock gravity also satisfies the second law at first order.

\section{conclusion}\label{sec5}

In this paper, we investigated the linearized second law of the black hole in Lovelock gravity sourced by a conformally coupled scalar field when the perturbation matter fields satisfy the null energy condition. As we all know, the Wald entropy does not satisfy the linearized second law for the purely Lovelock gravity. To show the validity of linearized second law in scalar-hairy Lovelock gravity, we first considered the Wald entropy in the Vaidya-like black hole solution and showed that it does not satisfy the linearized second law for the case with higher curvature terms. Moreover, we also show that we cannot naively correct the entropy by adding the scalar field term of the Wald entropy to the JM entropy of the purely Lovelock gravity to get a valid linearized second law. Then, by rescaling the metric $\tilde{g}_{ab}=\f^2 g_{ab}$, we can see that the action of the scalar field can be written as a Lovelock term with the metric $\tilde{g}_{ab}$. Using this property, by analogy with the JM entropy of the purely Lovelock gravity, we introduce a new formula of the entropy after adding the scalar field and showed that this JM entropy increases in Vaidya-like black hole solution for the scalar-hairy Lovelock gravity under first-order approximation. Moreover, we showed that different from the entropy in $F($Riemann$)$ gravity obtained in \cite{Wall}, here the difference between the JM entropy and Wald entropy contains the correction from the scalar field.

\section*{acknowledge}
 This research was supported by NSFC Grants No. 11775022 and 11873044.


\begin{thebibliography}{100}
\bibitem{A1}
  J.~D.~Bekenstein, ``Black holes and entropy,'' Phys.\ Rev.\ D {\bf 7}, 2333 (1973).
\bibitem{A2}
  J.~D.~Bekenstein,``Black holes and the second law,''  Lett.\ Nuovo Cim.\  {\bf 4}, 737 (1972).
\bibitem{A3}
  J.~D.~Bekenstein, ``Generalized second law of thermodynamics in black hole physics,'' Phys.\ Rev.\ D {\bf 9}, 3292 (1974).
\bibitem{A4}
  S.~W.~Hawking, ``Particle Creation by Black Holes,'' Commun.\ Math.\ Phys.\  {\bf 43}, 199 (1975).
\bibitem{A5}
  P.~C.~W.~Davies, ``Scalar particle production in Schwarzschild and Rindler metrics,'' J.\ Phys.\ A {\bf 8}, 609 (1975).
\bibitem{A6}
  W.~G.~Unruh, ``Notes on black hole evaporation,'' Phys.\ Rev.\ D {\bf 14}, 870 (1976).
\bibitem{A7}
 S.~W.~Hawking, ``Gravitational Radiation from Colliding Black Holes,'' Phys.\ Rev.\ Lett. {\bf 26}, 1344 (1971).
\bibitem{A8}
 M.~Bardeen, B.~Carter, and S.~W.~Hawking, ``The four laws of black hole mechanics,'' Commun. Math. Phys. {\bf 31}, 161 (1973).
\bibitem{A9}
 J.~D.~Bekenstein, ``Generalized second law of thermodynamics in black hole physics,'' Phys. Rev. D {\bf 9}, 3292 (1974).
\bibitem{A10}
 A.~C.~Wall, ``Proof of the generalized second law for rapidly evolving Rindler horizons,'' Phys.\ Rev.\ D {\bf 82}, 124019 (2010).
\bibitem{A11}
 A.~C.~Wall, ``Proof of the generalized second law for rapidly changing fields and arbitrary horizon slices,'' Phys. Rev. D {\bf 85}, 104049 (2012).
\bibitem{A12}
 A.~C.~Wall, ``Ten proofs of the generalized second law,'' J. High Energy Phys. {\bf 06} (2009) 021.
\bibitem{A13}
 B.~Zwiebach, ``Curvature Squared Terms and String Theories,'' Phys.\ Lett.\  {\bf 156B}, 315 (1985).
\bibitem{A14}
 D.~J.~Gross and E.~Witten, ``Superstring Modifications of Einstein's Equations,'' Nucl.\ Phys.\ B {\bf 277}, 1 (1986).
\bibitem{A15}
 A.~Sen, ``Black Hole Entropy Function, Attractors and Precision Counting of Microstates,'' Gen.\ Rel.\ Grav.\  {\bf 40}, 2249 (2008).
\bibitem{A16}
A.~Dabholkar and S.~Nampuri, ``Quantum black holes,'' Lect.\ Notes Phys.\  {\bf 851}, 165 (2012).
\bibitem{A17}
R.~M.~Wald, ``BlackholeentropyistheNoethercharge,'' Phys.\ Rev. D\ {\bf 48}, R3427 (1993).
\bibitem{A18}
V.~Iyer and R.~M.~Wald, ``Some properties of Noether charge and a proposal for dynamical black hole,'' Phys.\ Rev.\ D {\bf 50}, 846 (1994).
\bibitem{A19}
 S.~Gao and R.~M.~Wald, ''``Physical process version'' of the first law and the generalized second law for charged and rotating black holes,'' Phys. Rev. D {\bf 64}, 084020 (2001).
\bibitem{A20}
T.~Jacobson and R.~Parentani, ``Horizon entropy,'' Found. Phys. {\bf 33}, 323 (2003).
\bibitem{A21}
T.~Jacobson, G.~Kang, and R.~C.~Myers, ``Increase of black hole entropy in higher curvature gravity,'' Phys. Rev. D {\bf 52}, 3518 (1995).
\bibitem{A22}
T.~Jacobson, G.~Kang, and R.~C.~Myers, ``On black hole entropy,'' Phys. Rev. D {\bf 49}, 6587 (1994).
\bibitem{A23}
L.~H.~Ford and T.~A.~Roman, ``Classical scalar fields and violations of the second law,'' Phys. Rev. D {\bf 64}, 024023 (2001).
\bibitem{A24}
S.~Sarkar and A.~C.~Wall, ``Second law violations in Lovelock gravity for black hole mergers,`` Phys. Rev. D {\bf 83}, 124048 (2011).
\bibitem{Bhattacharjee:2015yaa}
  S.~Bhattacharjee, S.~Sarkar and A.~C.~Wall, ``Holographic entropy increases in quadratic curvature gravity,''
  Phys.\ Rev.\ D {\bf 92}, 064006 (2015).
\bibitem{A25}
 A.~Chatterjee and S.~Sarkar, ``Physical Process First Law and Increase of Horizon Entropy for Black Holes in Einstein Gauss-Bonnet Gravity,'' Phys. Rev. Lett. {\bf 108}, 091301 (2012).
\bibitem{A26}
 S.~Kolekar, T.~Padmanabhan, and S.~Sarkar, ``Entropy increase during physical processes for black holes in Lanczos-Lovelock gravity,'' Phys.\ Rev.\ D \ {\bf86}, 021501 (2012).
\bibitem{A27}
 S.~Sarkar and A.~C.~Wall, ``Generalized second law at linear order for actions that are functions of Lovelock densities,'' Phys. Rev. D {\bf 88}, 044017 (2013).
\bibitem{Wall}
 A.~C.~Wall, ``A Second Law for Higher Curvature Gravity,'' Int.\ J.\ Mod.\ Phys.\ D {\bf 24}, no. 12, 1544014 (2015).
\bibitem{B1}
J.~Oliva and S.~Ray, ``Conformal couplings of a scalar field to higher curvature terms,'' Class. Quant. Grav. {\bf 29}, 205008 (2012).
\bibitem{B2}
G.~Giribet, M.~Leoni, J.~Oliva and S.~Ray, ``Hairy black holes sourced by a conformally coupled scalar field in D dimensions,'' Phys. Rev. D {\bf 89}, 085040 (2014).
\bibitem{B3}
G.~Giribet, A.~Goya and J.~Oliva, ``Different phases of hairy black holes in AdS5 space,'' Phys. Rev. D {\bf91}, 045031 (2015).
\bibitem{B4}
M.~Galante, G.~Giribet, A.~Goya and J.~Oliva, ``Chemical potential driven phase transition of black holes in antide Sitter space,'' Phys. Rev. D {\bf92}, 104039 (2015).
\bibitem{Hennigar:2016ekz}
  R.~A.~Hennigar, E.~Tjoa and R.~B.~Mann, ``Thermodynamics of hairy black holes in Lovelock gravity,'' J. High Energy Phys. {\bf 1702} (2017) 070 (2017).
\bibitem{RPenrose}
R.~Penrose, ``Gravitational collapse: The role of general relativity,''
  Riv.\ Nuovo Cim.\  {\bf 1}, 252 (1969).
\end{thebibliography}
\end{document}